\RequirePackage{fix-cm}
\documentclass[twocolumn]{svjour3}          
\smartqed  
\usepackage{graphicx}
\usepackage{amsfonts}      
\usepackage{amsmath}      
\newcommand{\N}{\mathbb{N}}

\def\nerr#1#2 {{+{#1}}-{#2}}

\newcommand{\eqs}[2]{Equations~(\ref{#1}) and (\ref{#2})}

\newcommand{\bc}{\begin{center}}
\newcommand{\ec}{\end{center}}
\newcommand{\be}{\begin{equation}}
\newcommand{\ee}{\end{equation}}
\newcommand{\ba}{\begin{eqnarray}}
\newcommand{\ea}{\end{eqnarray}}
\begin{document}

\title{Difficulties of Quantitative Tests of the Kerr-Hypothesis with X-Ray Observations of Mass Accreting Black Holes}
 
\titlerunning{X-Ray Tests of the Kerr Hypothesis}        

\author{Henric Krawczynski}


\institute{H. Krawczynski \at
              Physics Department and the McDonnell Center for the Space Sciences 
              Washington University in Saint Louis
              1 Brookings Dr., CB 1105
              Saint Louis, MO 63130
               \\
              Tel.: +314-935-8553\\
              \email{krawcz@wustl.edu} 
}
\date{Printed: \today}
\maketitle

\begin{abstract}
X-ray studies of stellar mass black holes 
in X-ray binaries and mass-accreting supermassive 
black holes in Active Galactic Nuclei have achieved 
a high degree of maturity and have delivered detailed information 
about the astrophysical sources and the physics of black hole accretion. 
In this article, I review recent progress made towards using the X-ray observations for 
testing the ``Kerr hypothesis'' that the background spacetimes of all astrophysical 
quasi-stationary black holes are described by the Kerr metric. 
Although the observations have indeed 
revealed clear evidence for relativistic effects in strong-field gravity, 
quantitative tests of the Kerr hypothesis still struggle with theoretical and practical difficulties.
In this article, I describe several recently introduced test metrics and 
review the status of constraining the background spacetimes of 
mass accreting stellar mass and supermassive black holes with these test metrics. 
The main conclusion of the discussion is that astrophysical uncertainties 
are large compared to the rather small observational differences between 
the Kerr and non-Kerr metrics precluding quantitative constraints 
on deviations from the Kerr metric at this point in time. 
I conclude with discussing future progress enabled 
by more detailed numerical simulations and by future X-ray spectroscopy, 
timing, polarimetry, and interferometry missions. 
\end{abstract}

\section{Introduction}
\label{intro}
General Relativity's (GR's) no-hair theorem states that the Kerr \cite{Kerr:63,Teuk:15} 
and Kerr-Newman \cite{Newm:65,Newm:14} families of background spacetimes are the 
only stationary, axially symmetric, and asymptotically flat vacuum (Kerr) and 
electro-vacuum (Kerr-Newman) solutions of the Einstein equations that have an 
event horizon and neither singularities nor closed timelike curves in the outer domain
\cite{Isra:67}. As astrophysical black holes are thought to be largely electrically 
neutral \cite{Blan:77,Thor:86},  the no-hair theorem implies the ``Kerr hypothesis'' 
that if GR and the assumptions made for deriving the no-hair theorem are valid, 
the background spacetimes of all quasi-stationary astrophysical black holes 
can be described by the Kerr family of metrics.  If the Kerr hypothesis holds, astrophysical 
black holes are as simple as elementary particles. Of course, tests of the Kerr hypothesis 
only depend on the spacetime geometry of black holes, and do not test the dynamical 
aspects of the underlying theory.

The present article follows several excellent reviews of tests of GR in the weak and strong gravity 
regimes and tests of the Kerr hypothesis. 
In his review ``The Confrontation between General Relativity and Experiment'', 
Will posits that the tests of Einstein's equivalence principle (the trajectories of freely falling bodies 
do not depend on their structure and composition plus local Lorentz and position invariance) 
with exquisite accuracy strongly argue for gravity being described by a metric theory \cite{Will:14}. After reviewing theoretical frameworks for testing 
GR such as the Parametrized Post Newtonian (PPN) formalism, he gives a broad overview of
alternative theories of gravity including scalar tensor theories, $f(R)$ theories, vector tensor theories, 
tensor-vector-scalar theories, quadratic gravity, and massive graviton theories. He summarizes the state 
of the art of GR's validation based on solar system and stellar system observations, and outlines the potential
of gravitational wave detections. 

In the review article ``Probes and Tests of Strong-Field Gravity with Observations in the 
Electromagnetic Spectrum'' 
Psaltis argues for the need of testing GR in the strong-gravity regime \cite{Psal:08}. 
He characterizes the strength of the gravitational field of an object of mass $M$ at a distance $r$ 
with the depth of the potential well $\varepsilon=GM/r\,c^2$ and the spacetime curvature 
$\xi=GM/r^3 c^2$. Observations of phenomena close to event horizons of rapidly spinning stellar mass black holes test GR at $\sim$10 orders of magnitude larger $\varepsilon$-values and $\sim$5 orders of magnitude larger $\xi$-values than solar system tests and observations of the Hulse-Taylor binary pulsar.
 Psaltis describes different observational channels for GR tests, 
 i.e.\ electromagnetic observations of inspiralling compact binaries, 
 Very Long Baseline Interferometric observations of the shadows of the black holes
in Sgr A$^*$ and M~87, X-ray evidence for the existence of stellar mass black holes, 
lines in the X-ray energy spectra of black holes and neutron stars, 
black hole size constraints from X-ray timing observations, 
and the analysis of the quasi-periodic oscillations of the X-ray fluxes from mass accreting 
neutron stars and black holes.  

The three reviews ``Gravitational-Wave Tests of General Relativity with Ground-Based Detectors 
and Pulsar-Timing Arrays'' by Yunes \& Siemens 
\cite{Yune:13},  
``Testing general relativity with present and future astrophysical observations'' by 
Berti et al. \cite{Bert:15}, and
``Testing General Relativity with Low-Frequency, Space-Based Gravitational-Wave Detectors'' 
by Gair, Vallisneri, Larson, and Baker 
\cite{Gair:16} follow up on Will's review. 
The authors give an expanded review of alternative theories of gravity and 
discuss the existence and properties of compact objects and gravitational waves in these theories.
The interested reader is referred to Tables 1-3 of Berti et al.'s paper for summaries of
alternative theories of gravity and how they relate to the assumptions of Lovelock's theorem (Table 1),
and the properties of black holes (Table 2) and neutron stars (Table 3) in these theories. 
The three reviews summarize strong-field tests of GR based on gravitational wave detections. 
Yunes \& Siemens formulate criteria which make an alternative theory of gravity 
a well motivated and useful tool for strong-field tests of GR: a theory should be motivated 
from a fundamental physics standpoint, predictive for a range of initial value data, 
consistent with all current tests of GR, and should make strong-field predictions deviating from GR.  
They conclude that most alternative theories of gravity fail to meet one or several of these criteria 
with the exception of scalar-tensor theories with spontaneous scalarization which meet
the first three criteria and possibly the fourth criterion.  

The two reviews ``Testing the no-hair theorem with observations of black holes in the electromagnetic spectrum'' 
by Johannsen \cite{Joha:16} 
and  ``Testing black hole candidates with electromagnetic radiation'' 
by Bambi, Jiang, and Steiner \cite{Bamb:17}
are closely related to this paper. Both papers discuss tests of the Kerr hypothesis using
parametric test metrics, and present predicted observational results for a range of 
different parameter values. Johannsen's review includes a detailed description of the status of
testing the Kerr hypothesis with radio interferometric observations of the black hole shadows of
Sgr A$^*$ and M~87, and stars orbiting Sgr A$^*$. Compared to these earlier reviews,
I will emphasize in this paper that X-ray tests of the Kerr hypothesis are currently of limited value
owing to large systematic uncertainties stemming from astrophysical uncertainties and from
modeling the systems with restrictive assumptions and limited fidelity.

Following Will's rationale, we will focus the discussion in this paper on tests of GR versus 
other metric theories of gravity in which the gravitational effects are fully described by the metric,
the law of energy-momentum conservation holds  (i.e. the covariant
derivative of the energy-momentum tensor vanishes: ${\bf \nabla}\cdot{\bf T}=0$), and
photons and massive particle follow geodesics of the metric. 
In the case of theories that can be derived from extremizing an action, the law of 
energy-momentum conservation can be derived from the requirement of the 
diffeomorphism invariance of the action as long as there is a separation of the 
matter and field actions (see \cite{Caro:03} and also \cite{Fock:64,Thor:71,Soti:08}).
Electromagnetic tests of the Kerr hypothesis then involve the comparison of predictions 
derived for the Kerr spacetime and alternative spacetimes. The latter may 
result from alternative theories of gravity, or may be of purely phenomenological nature. 
In both cases, the metrics depend not only on a mass and an angular momentum parameter, 
but also on additional deviation parameters.
All the metrics considered in the following include the Kerr family of metrics for certain values of the
deviation parameters (i.e. 0 for additive terms, or 1 for multiplicative terms).
Unfortunately, strong gravity tests lack a canonical parameterization similar to the PPN parameterization
used for testing GR in the weak-field and slow-motion limit \cite{Will:14,Pois:14}. 
For many alternative theories of gravity it is not known if black hole solutions exist. If they do,
the metrics have only been derived in the limit of low spin values with the exception of
Einstein-Gauss-Bonnet-Dilaton gravity for which a high-spin metric has been derived 
numerically \cite{Klei:11}. 

The rest of the paper is structured as follows. I start with a brief description of some of the key properties
of accretion disks and coronas of mass accreting stellar mass and supermassive black holes in Sect.\ \ref{Xintro}.
After introducing several test metrics in Sect.\ \ref{Xmetric}, I discuss some of the physical properties of
the spacetimes they describe in Sect. \ref{Xphys}. Subsequently, I discuss the two main methods 
that are being used to constrain the properties of the background spacetimes of astrophysical 
black holes in Sect.\ \ref{Xobs} and summarize the results obtained for stellar mass black holes. 
The results confirm key GR predictions, i.e. that matter can orbit a black hole close to its event horizon
emitting radiation with large gravitational and Doppler frequency shifts in agreement with the GR predictions.    
However, systematic uncertainties are still too large to allow for robust tests of the Kerr metric against 
other metrics. I summarize my conclusions in Sect.\ \ref{disc}, and discuss avenues for future progress.

Most of the reviews mentioned above were written before the detection of gravitational 
waves from the binary stellar mass black hole mergers GW150914 \cite{Abbo:16a}, 
GW151226 \cite{Abbo:16b}, GW170104 \cite{Abbo:17a}, GW170608 \cite{Abbo:17b}, 
and GW170814 \cite{Abbo:17c} and the binary neutron star merger GW170817 \cite{Abbo:17d}.
The discoveries by LIGO and now also by VIRGO and the radio to gamma-ray observations of the electromagnetic
counterpart of GW170817 have already led to important tests of GR and fundamental physics.
The results include the test of GR's prediction that gravitational waves and gamma-rays propagate 
with the same speed with an accuracy of a few parts in 10$^{15}$ \cite{LIGO:17}, 
the confirmation that the spacetime curvature generated by the Milky Way affects gravitational waves 
and gamma-rays in the same way within an accuracy of a few parts in 10$^{6}$ \cite{LIGO:17}, 
and the test of GR's prediction of two spin-2 tensor polarizations of 
gravitational waves \cite{Abbo:17c}. The gravitational wave event GW150914 was used for parametric tests,
i.e. constraints on higher-order post-Newtonian parameters and on GR deviation parameters 
of parametrized waveforms \cite{Abbo:16aa}.   
The results show the power of opening up a new observational window, 
and the fundamental physics insights that can be gained from observing dynamical gravity in action. 

In the following we use the definition of the black hole region of an asymptotically flat spacetime 
being the region from which no future-pointing null geodesic can reach future null infinity. 
The event horizon (a null hypersurface in four dimensional spacetime) is defined 
as the boundary of this region.
Throughout the paper, we use geometric units ($G$=$c$=1) and express all distances
in units of the gravitational radius $r_{\rm g}=GM/c^2$ with $M$ being the black hole
mass. Denoting the angular momentum by $J$, the spin parameter is given by 
$a=J/c r_{\rm g} M=c J/ G M^2$. In units of $M$, the spin parameter $a$ can range from -1 to +1.
\section{Black Hole Accretion Disks and Coronas}  
\label{Xintro}
The theory of geometrically thin, optically thick accretion disks is based on the papers by 
Shakura \& Sunyaev (1973) \cite{Shak:73}, Novikov \& Thorne \cite{Novi:73}, and Page \& Thorne \cite{Page:74}. 
The matter is assumed to orbit the black hole on near-Keplerian orbits in the equatorial plane of the spacetime.
Turbulent viscosity described by the $\alpha$-parameter transports angular momentum outwards
enabling matter to flow inwards. The accreting matter is assumed to move from circular to circular orbit,
locally emitting all the excess gravitational energy as it sinks towards the black hole. 
Assuming that the matter plunges into the black hole once it reaches the Innermost Stable Circular Orbit
(ISCO) and the viscous torque vanishes at the ISCO, the conservation laws for mass, energy, 
and angular momentum fully determine the radial brightness distribution $F(r)$. 
The thin disk models are believed to hold for accretion luminosities 
$\dot{M} c^2$ (mass accretion rate times the speed of light squared) 
between a few percent 
and several ten percent of the Eddington Luminosity  
$L_{\rm Edd}=4\pi G M  c /\kappa_{\rm es}$ (with the electron 
scattering opacity being $\kappa_{\rm es}\approx 0.4\, {\rm cm^2 g^{-1}}$).
In this regime, the gravitational potential energy of the accreted mass is believed 
to be efficiently converted into radiation with an accretion efficiency 
$\eta\equiv L/\dot{M}c^2$ on the order of $\sim$10\% ($L$ being the bolometric luminosity of the emission). 
At extremely low and extremely high accretion rates, the 
accretion efficiency $\eta$  is likely to be much lower as the matter 
is either too tenuous to radiate efficiently 
for $L\ll L_{\rm Edd}$ or advects diffusively trapped photons alongside 
the matter into the black hole for $L\ge L_{\rm Edd}$. The magnetorotational instability (MRI) 
is believed to be the prime source of the viscosity driving the accretion \cite{Balb:98}.  
Tests of the Kerr hypothesis are usually based on systems believed to accrete via
geometrically thin and optically thick accretion disks. 
Descriptions of accretion flows in other regimes can be found in 
\cite{Abra:13,Feng:14}.

General relativistic magnetohydrodynamical (GRMHD) and general relativistic radiation 
magnetohydrodynamic (GRRMHD) simulations make it now possible to perform ab-initio simulations of accretion flows. 
GRMHD simulations have largely confirmed the results of the analytical thin disk 
theory, showing that the radial brightness distribution matches that of the analytical models to good
approximation, and the emission from within $r_{\rm ISCO}$ accounts for only a few percent 
of the total disk luminosity \cite{Nobl:11,Penn:12}. 
Ongoing work includes studies of the transport of magnetic field and 
angular momentum (e.g.\ \cite{Mars:17,Fouc:17}), line blanketing and 
thermal instabilities (e.g. \cite{Jian:16,Mish:16}), radiation transport (e.g.\ \cite{Sado:16}),  
the Comptonization of photons (e.g. \cite{Nara:16}), and
the impact of different ion and electron temperatures on the accretion in the
low-luminosity regime (e.g.\ \cite{Ryan:17}). Some of the current GRRMHD developments
parallel similar earlier developments in stellar radiation magnetohydrodynamics, 
see \cite{Cast:04} for an introduction and review.

For thin disks and a given accretion rate $\dot{M}$ in units of the Eddington luminosity $L_{\rm Edd}$, 
the radiated luminosity scales as $L_{\rm bol}\propto M$ (see Equ.\ (\ref{EPT2}) below). 
The energy flux $F$ per proper disk area scales as 
$F\propto L_{\rm bol}/r_{\rm g}^{\,2}\propto M^{-1}$, 
and the energy scale of the emitted photons scales with the 
photospheric temperature $T\propto F^{1/4}\propto M^{-1/4}$ (see Equ.\ (\ref{temp}) below). 
Whereas the thermal disk emission of stellar mass black holes has keV energies and can readily 
be observed with X-ray telescopes, that of AGNs is in the blue and UV bands and is often masked
by the emission from other components of the accretion flow.
The good agreement of Shakura, Sunyaev, Novikorn, and Thorne's analytical thin disk model 
with the results of numerical simulations  explain its continued use more than four decades 
after its invention. The model can explain the observed X-ray energy spectra of the thermal
state of accreting stellar mass black holes such as  LMC X-3 \cite{Stei:14}, and the blue bumps 
in the spectral energy distributions (SEDs) of active galactic nuclei (AGNs) \cite{Elvi:94}.
Some observations are not yet unambiguously explained, or seem to be in tension with thin disk theory. 
For example, optical microlensing observations of gravitationally lensed quasars \cite{Morg:10},
photometric quasar variability studies \cite{Jian:17}, and optical reverberation observations \cite{Edel:15}
all indicate that optical accretion disks are by a factor of $\sim 2$ larger than predicted 
by thin disk theory (see \cite{Hall:17} for a possible explanation).

\begin{figure}[t]
\begin{center}
\includegraphics[width=0.45\textwidth]{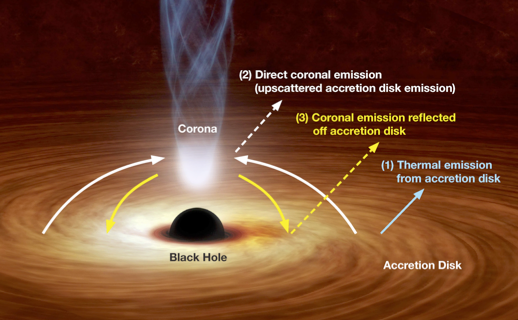}
\end{center}
\caption{\label{jpl} Artist's impression showing (1) the thermal emission from the accretion disk, 
(2) the coronal emission, and (3) the reflected and reprocessed coronal emission (courtesy of JPL/NASA).    }
\end{figure}

For both types of black holes, mass accreting stellar mass and supermassive black holes, 
a corona of hot plasma is believed to emit X-rays with a power law energy spectrum.
Some of the X-rays irradiate the accretion disk prompting the emission of 
Fe K$\alpha$ fluorescent emission at plasma frame energies around 6.4 keV 
and reflected Compton hump emission (Fig. \ref{jpl}, 
see the {\it NuSTAR} results for Cyg X-1 for exemplary energy spectra  \cite{Toms:14}).
The shape and location of the corona is still a matter of debate (see e.g.\ \cite{Gilf:14}).
For AGNs, the shape of the relativistically broadened Fe K$\alpha$ line  \cite{Wilm:01,Fabi:09,Chia:15} 
and time lags inferred from the Fe K$\alpha$--power law continuum cross correlation function \cite{Uttl:14}
hint at very compact coronas very close (within a few $r_{\rm g}$) to the black holes.
Additional evidence for compact coronas comes from the amplitude of the brightness fluctuations 
of the X-ray emission from gravitationally microlensed quasars (see Sect.\ \ref{disc}).

X-ray astronomers often assume that the corona has a negligible spatial extent and is located 
on the rotation axis of the black hole and thus irradiates the accretion disk in a lamppost configuration 
\cite{Matt:91}. GRMHD simulations indicate that such a configuration 
may arise from hot magnetized plasma rising buoyantly towards the underdense polar regions  
above and below the black hole \cite{Schn:13}.  Alternatively, the corona may be related 
to the launching site of a jet. Symmetry considerations argue for the existence of 
two of such coronas, one in each hemisphere of the black hole.
For mass accreting supermassive black holes, there is some tension between the corona being 
sufficiently small to account for the results from gravitational lensing, spectroscopic and 
timing studies, and being sufficiently large to explain the observed X-ray fluxes via the
Comptonization of accretion disk photons passing through the corona \cite{Dovc:16}.  
Current state-of-the-art models usually assume that the photosphere of the entire accretion disk
can be described with a constant ionization parameter and a constant density of the accretion disk photosphere \cite{Ross:05,Garc:13}. Eventually, more detailed numerical simulations should enable us
to develop fitting models that account for the radial dependence of the ionization state 
and the density of the emitting plasma (see \cite{Garc:16,Toms:18}).
\section{Black Hole Test Metrics}  
\label{Xmetric}
Using the coordinates $x^{\mu}=(t,r,\theta,\phi)$, any stationary, axially symmetric and
asymptotically flat metric can be brought into standard form:
\begin{equation}
ds^2 
=\,-e^{-2\nu_0}\,dt^2+e^{2\psi} (d\phi-\omega dt)^2+e^{2\mu_1}dr^2+e^{2\mu_2}d\theta^2 
\label{metric}
\end{equation}
with $\nu_0$, $\psi$, $\omega$, $\mu_1$, and $\mu_2$ being functions of $r$ and $\theta$ only \cite{Chan:83}.
Assuming invariance under simultaneous inversion of $t$ and $\phi$,  $\mu_1= \mu_2$ 
already covers all possible metrics, 
but not requiring this equality provides for advantageous gauge choices. 
Setting $M=1$ for convenience, the Kerr metric is given in Boyer Lindquist coordinates by \cite{Boye:67}:
\[
e^{-2\nu_0}=\Sigma\Delta/A\,,\,
e^{2\psi}={\rm sin^2}\theta A/\Sigma\,
\]\[
e^{2\mu_1}=\Sigma/\Delta\,,\,
e^{2\mu_2}=\Sigma
\]\[
\omega=2ar/A\,,
\Delta=r^2-2r+a^2\,,
\]
\begin{equation}
\vspace*{-0.5cm}
\Sigma=r^2+a^2{\rm cos}^2\theta\,,\,
A=(r^2+a^2)^2-a^2\Delta\,{\rm sin}^2\theta.
\label{kerr}
\end{equation}

The desire to quantitatively test the Kerr hypothesis has led to the development of 
a number of parametric test metrics. For the purpose of the this paper, a test metric 
{\it m} is a map of $N$ real parameters $(a_1, a_2, ... a_N)$ to stationary, axially symmetric,  
asymptotically flat black hole metrics $g(a_1, a_2, ... a_N)$. A particular metric $g$ is a 
black hole metric if the spacetime possesses an event horizon and no ``pathologies'', 
i.e. neither curvature singularities at and outside of the event horizon, 
nor violations of the Lorentzian signature det$(g) < 0$ or closed timelike curves outside 
of the event horizon. The metrics $g$ may or may not be solutions of a field equation.   
A test metric is particular useful if it satisfies a number of criteria:
\begin{description}
\item [{\it C1}:] The range of the map $R(m)$ includes the Kerr family of metrics. 
\item [{\it C2}:] $R(m)$ includes at least one valid black hole spacetime (i.e. a black hole
spacetime without pathologies in the exterior domain) which is physically distinct from any Kerr spacetime.
\item [{\it C3}:] $R(m)$ covers a wide range of physically different black hole spacetimes.
\item [{\it C4}:] ${R(m)}$ includes the black hole metrics from one or several alternative theories of 
gravity such that observational constraints on the parameters $a_1, a_2, ... a_N$ 
can be translated into constraints on the parameters of the alternative theories of gravity.
\item [{\it C5}:] The individual parameters of the test metric, or combinations of these parameters, 
can be identified with certain physical properties of the spacetimes.
\end{description}

In the remainder of this section, I discuss these criteria for a few metrics from the literature:
\begin{description}
\item [{\it m1}:]  Pani et al. (2011) derived slowly spinning black hole metrics for theories of gravity with 
an Einstein-Hilbert action augmented by quadratic and algebraic curvature invariants coupling 
to a single scalar field \cite{Pani:11}. 
\item [{\it m2}:] Aliev and G{\"u}mr{\"u}k{\c c}{\"u}o{\v g}lu (2005) argue that GR's Kerr-Newman
metric describes a black hole on a 3-brane in the Randall-Sundrum braneworld \cite{Alie:05}.
The Kerr-Newman charge parameter is interpreted to describe a ``tidal charge''. 
\item [{\it m3}:] Ghasemi-Nodehi \& Bambi (2016) introduce 11 parameters modifying every occurrence of 
the mass parameter $M$ and the spin parameter $a$ in the Kerr metric, allowing them to modify how these
two properties couple to the spacetime curvature \cite{Ghas:16}.
\item [{\it m4}:] The Geroch-Hansen mass multipole moments $M_{\rm l}$ and current multipole  moments $S_{\rm l}$ ($l\in {\N}_0$) of the Kerr metric depend only on two parameters  $M$ and $a$: $M_{\rm l}^{\rm K}+iS_l^{\rm K}=M(ia)^l$ \cite{Gero:70,Hans:74,Fodo:89}.  Glampedakis and Babak (2006) 
introduce a test metric by modifying the quadrupole moment 
with the help of a dimensionless deviation parameter $\varepsilon$: 
$M_{\rm 2}=M_{\rm 2}^{\rm K}-\varepsilon M^3$ 
and neglecting deviations in all higher moments \cite{Glam:06}.  

\item [{\it m5}:] Johannsen \& Psaltis (2011) derive a test metric (called {\it m5a} in the following)
by applying the Newman-Janis algorithm to a Schwarzschild metric with the $(t,t)$ and $(r,r)$
components modified by a multiplicative factor of $(1+h(r))$  and expanding the
real function $h(r)$ in powers of $1/r$ \cite{Joha:11}. Cardoso, Pani \& Rico (2014) 
derive a generalized version of this test metric (called {\it m5b} in the following) 
using a seed metric with different modifiers for the $(t,t)$ and $(r,r)$ elements 
of the Schwarzschild metric, doubling the number of free parameters \cite{Card:14}.

\item [{\it m6}:] Johannsen (2013,2016) constructs a test metric that leaves the Hamilton-Jacobi 
equations of a test particle separable \cite{Joha:13,Joha:16} (see also the earlier work of
 \cite{Vige:11}). Johannsen's metric allows for  three constants of motion, two associated 
with the Killing vectors of the temporal and axial symmetries, plus one ``Carter constant'' following from the separability of the Hamilton-Jacobi equations. 
The  metric uses one parameter $\beta$ and four functions $A_1(r)$, $A_2(r)$, $A_5(r)$, and $f(r)$ 
to modify the $(t,t)$, $(t,\phi)$, $(r,r)$, $(\theta,\theta)$ and $(\phi,\phi)$ components of the Kerr metric. 
Expanding $A_1(r)$, $A_2(r)$, $A_5(r)$ around $1$ and
$f(r)$ around 0 in powers of $1/r$ leads to a set of expansion coefficients called 
$\alpha_{1n}$, $\alpha_{2n}$, $\alpha_{5n}$, and $\epsilon_n$, respectively, with $n$ designating the power of $1/r$. 
\end{description}
The test metrics {\it m4}, {\it m5}, and {\it m6} use power series in $1/r$ which should converge 
for $r/M\gg 1$, but which do not necessarily converge in the regime $r/M\sim 1$ which
is of particularly interest for X-ray tests of the Kerr hypothesis. 
Konoplya, Rezzolla and Zhidenko performed detailed studies of the convergence of 
various parameterizations of generally axially symmetric test metrics. 
They find that a continued-fraction expansion in terms of a compactified radial 
coordinate and a Taylor expansion in terms of the cosine of the polar angle shows 
excellent convergence -- at least in the equatorial plane \cite{Kono:16}.

A test metric with a range including the Kerr family of metrics (criterion {\it C1}) can be used to constrain deviations from GR. All metrics {\it m1}-{\it m6} satisfy this criterion.  
Criterion {\it C2} requires that the range of the test metric includes one or more non-Kerr spacetimes. 
In Section \ref{Xphys} we will discuss a prescription that can be used to show that a test metric
indeed produces spacetimes that a physically different from any Kerr spacetime 
(accounting for the gauge freedom). In addition, criterion {\it C2} requires that the metrics 
are regular in the exterior domain, assuring that numerical simulations can be run for these
spacetimes.  We know that the test metrics  {\it m2} and {\it m6} satisfy this criterion, 
and that the test metrics {\it m4} and {\it m5a} do not.
The non-Kerr metrics of the test metric {\it m4} are known to  exhibit pathologies 
in the exterior domain, and the non-Kerr metrics of the test metric {\it m5a} (and {\it m5b}) 
exhibit curvature singularities at the event horizon \cite{Joha:13b,Card:14}. 
To my knowledge, the regularity of the metrics {\it m1} and {\it m3} has not yet 
been studied so far. Even if a test metric does not satisfy criterion {\it C2}, it can still be used 
as a tool to scrutinize observational data for deviations from the Kerr spacetimes 
by neglecting any radiation entering or coming from the affected portions 
of the spacetimes, see e.g.\ \cite{Joha:11,Kraw:12,Hoor:16,Joha:16,Bamb:17}.

Criteria {\it C3} and {\it C4} concern the richness of the spacetimes of the test metric. 
One possible way of showing that a test metric covers a non-trivial set of 
physically different spacetimes (criterion {\it C3}) is to show that it includes 
known black hole solutions from one or several alternative 
theories of gravity (criterion {\it C4}). Cardoso, Pani and Rico (2014) show that the test metric 
{\it m5b} covers the non-spinning black holes of the Einstein-Dilaton-Gauss-Bonnet gravity, 
but that neither {\it m5a} nor {\it m5b} cover the spinning counterparts \cite{Card:14}. 
The test metric {\it m6} covers the Kerr-Newman, Bardeen and modified gravity 
black hole spacetimes. For small values of the deviation parameter, the metric
captures the Einstein-Dilaton-Gauss-Bonnet and dynamical Chern-Simons black hole solutions
valid up to linear order in the spin parameter \cite{Joha:16}. 
Another way of assessing the richness of $R(m)$, is to evaluate the observational signatures
for all the parameters of the metric, see Sect.~\ref{Xphys}.      

Finally, one would wish that certain parameters of the test metric can be identified with certain physical 
properties of the spacetime (criterion {\it C5}). Ghasemi-Nodehi \& Bambi systematically explore the effect of the 11 
parameters of the test metric {\it m3} on the observed appearance of black hole shadows \cite{Ghas:16}.
Similarly, Table~3 of Johansen (2016) shows the impact of the parameters of the test metrics 
{\it m5a} and {\it m6} on certain types of observations. 

\section{Observational Signatures}  
\label{Xphys}
Combining the analytical models of the coronal or accretion disk emission described in Sect. \ref{Xintro} with 4-D ray tracing codes, predicted X-ray flux and polarization energy spectra, light curves, and images can be derived. Ray tracing schemes track photons from the observer to the black hole
\cite{Psal:12,Bamb:12}, from the accretion disk  or corona to the observer \cite{Kraw:12,Behe:17},
or make use of Cunningham's transfer function formalism \cite{Bamb:17}. 
Such codes have been used to study the impact of the background spacetime on Fe K$\alpha$ 
line shapes (e.g. \cite{Psal:12,Bamb:13,Joha:14,Hoor:16,Bamb:17}), 
continuum energy spectra (e.g.\ \cite{Kraw:12,Bamb:12,Joha:14,Hoor:16,Ayze:17}), 
polarization energy spectra (e.g. \cite{Kraw:12,Hoor:16}),  the shapes of black hole shadows 
(e.g.\ \cite{Kraw:12,Joha:13c,Hoor:16}),  and Fe K$\alpha$ reverberation signatures (e.g.\ \cite{Hoor:16}).
All these studies assume that the angular momentum vectors of the black holes and 
the accretion disks align.

\begin{figure}[t]
\begin{center}
\includegraphics[width=0.42\textwidth]{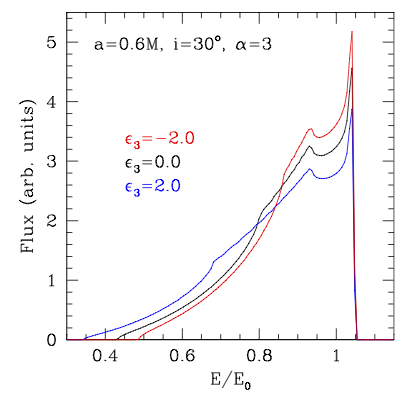}
\end{center}
\caption{\label{fj1} 
Simulated Fe K$\alpha$ energy spectra for the test metric {\it m5a} for $a = 0.6$, $i=30^{\circ}$
for several values of the parameter $\epsilon_3$ measuring the deviation from the Kerr metric. 
Reproduced from \cite{Joha:13a} with permission of the authors.}
\end{figure}

\begin{figure}[t]
\begin{center}
\includegraphics[width=0.427\textwidth]{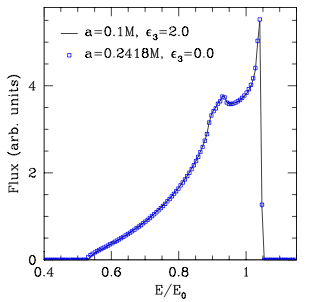}
\end{center}
\caption{\label{fj2} 
Same as Figure \ref{fj1} but for two combinations of spin parameter $a$ and 
deviation parameter $\epsilon_3$ giving the same $r_{\rm ISCO}$-values. 
The two metrics lead to almost identical energy spectra (black solid line and 
blue squares). Reproduced from \cite{Joha:13a} with permission of the author.
}
\end{figure}

\begin{figure*}[t]
\begin{center}
\includegraphics[width=0.9\textwidth]{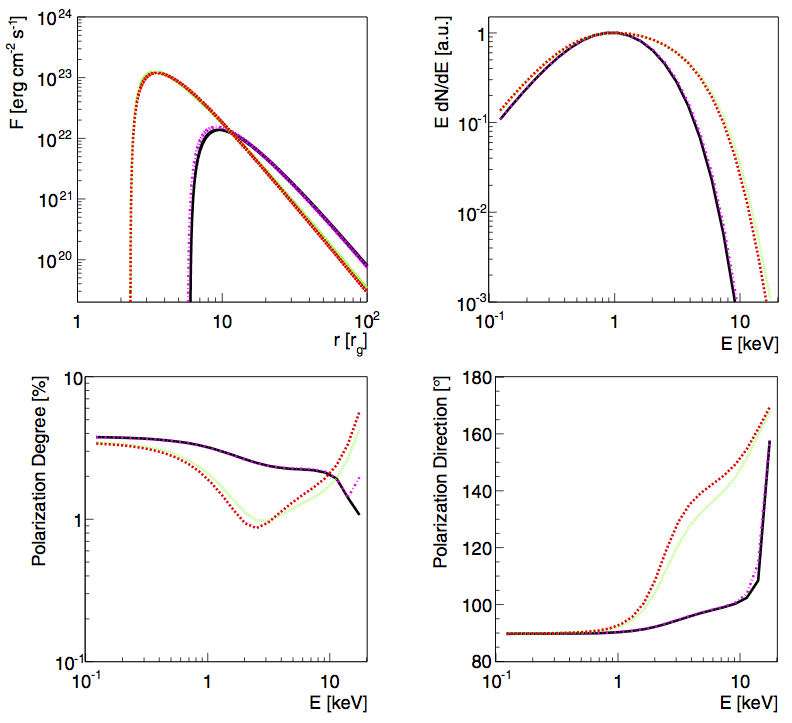}
\end{center}
\caption{\label{hk1} 
Accretion disk brightness in the plasma frame (top left), energy spectrum (top right), polarization fraction (bottom left), 
and polarization angle (bottom right) for two pairs of a Kerr metric and a non-Kerr metric giving similar 
$r_{\rm ISCO}$-values. The solid black line and the dashed-dotted magenta lines show the results for the Kerr metric
with $a=0$, $r_{\rm ISCO}=6$ and the {\it m5a} metric with $a = 0.5$, $\epsilon_3=-5$, $r_{\rm ISCO}=5.8$, respectively.
Similarly, the dotted green line and the dashed red line show the results for the Kerr metric with $a=0.9$ and
the {\it m5a} metric with $a=0.5$ and $\epsilon_3 = 6.3$, respectively, both giving $r_{\rm ISCO}=2.32$.
From \cite{Kraw:12}.}
\end{figure*}

\begin{figure}[t]
\begin{center}
\includegraphics[width=0.45\textwidth]{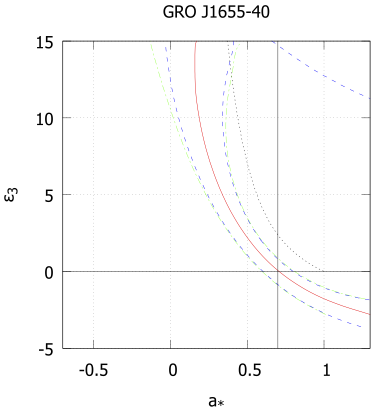}
\end{center}
\caption{\label{cb1} 
Constraints on the {\it m5a} parameters $a$ and $\epsilon_3$ derived from fitting the energy spectrum
of the thermal accretion disk emission predicted for the black hole candidate GRO J1655-40.
The red solid curve shows the best fit values. The blue dashed and green dashed-dotted lines show 
uncertainties resulting from the measurement errors. Reproduced from \cite{Kong:14} with permission of the authors.
}
\end{figure}

Figure \ref{fj1} shows that the parameter $\epsilon_3$ of the test metric {\it m5a} has 
a strong impact on the predicted Fe~K$\alpha$ inner disk reflection spectra (from \cite{Joha:13a}). 
However, in the same paper, the authors note that the predicted energy spectra agree surprisingly well 
when comparing Kerr and non-Kerr metrics giving identical $r_{\rm ISCO}$-values 
(Fig.~\ref{fj2}). We found a similar approximate degeneracy of the Kerr and non-Kerr metrics
in terms of the predicted observational signatures for the metrics {\it m1}, {\it m2}, {\it m4}, and {\it m5a}
as long as we compare metrics with identical or similar $r_{\rm ISCO}$-values.
Fig.~\ref{hk1} shows the predicted flux and polarization energy spectra of the thermal disk emission for
Kerr and non-Kerr models giving the same $r_{\rm ISCO}$-values (from \cite{Kraw:12}, see also 
\cite{Hoor:16}). The results demonstrate that the Kerr metric and the {\it m5a} metrics 
produce almost indistinguishable observational signatures for a considerable fraction of the 
{\it m5a} parameter space, including nominal non-Kerr metrics with a non-vanishing deviation
parameter $\epsilon_3$.  
Kong, Li, and Bambi (2014) explored the degeneracy of the Kerr and {\it m5a} metric with regards 
to observational signatures by fitting theoretical Fe K$\alpha$ inner disk reflection energy spectra 
with the energy spectra derived for {\it m5a} background spacetimes (Fig. \ref{cb1}). 
The results establish an equivalence class of metrics with $a$ and $\epsilon_3$-combinations 
mapping to near-identical observational results. Although it seems likely that the observationally 
degenerate metrics have similar $r_{\rm ISCO}$-values, the authors did not comment 
on this in their paper.

In summary, for large regions of the parameter space, the test metrics 
{\it m1}, {\it m2}, {\it m4}, and {\it m5a} predict almost identical observational 
signatures as suitably chosen Kerr metrics. The only exceptions seem to be spacetimes
with $r_{\rm ISCO}$-values larger than the maximum value 
$r_{\rm ISCO}=9$ for the Kerr spacetime.
The results demonstrate the difficulty of evaluating the criteria 
{\it C2} (the range of the metric includes one or more valid spacetimes 
which are physically distinct from the Kerr metric) and 
{\it C3} (the test metric covers a rich range of physically different spacetimes) in the presence of degeneracies 
between the parameters of the test metric, and the gauge freedom of metric theories.
One way of simplifying the analysis is to transition from using heavily degenerate metric 
parameters to parameters which are either related to observables or leave other observables unchanged. It is furthermore helpful to focus the discussion of the properties of a spacetime entirely 
on observables and to avoid using coordinates. In the remainder of this section, 
we demonstrate this procedure for the specific examples of the test metrics {\it m2} and {\it m6}. 
The test metric {\it m2} depends on the parameters $M$, $a$ and $\beta$. 
For the test metric {\it m6} we limit the discussion to the parameters $M$, $a$ and $\alpha_{22}$. 

Rather than performing full ray tracing simulations, we use basic analytical equations to derive some
pseudo-observables. We simplify the discussion of the physical properties of the spacetimes of the test metrics by using 
($M$, $P_{\rm ISCO}$, $a$) as the parameters characterizing a spacetime instead of 
($M$, $a$, $\beta$) or ($M$, $a$, $\epsilon_3$). The parameter $M$ describes for all metrics the
mass measured by a distant observer, $P_{\rm ISCO}$ is the orbital period of matter 
orbiting the black hole at the ISCO as measured by a distant observer in the 
asymptotically flat region of the spacetime. The parameter $a$ labels the one 
dimensional space of metrics with identical $M$ and $P_{\rm ISCO}$. 
The choice of comparing Kerr and non-Kerr metrics with the same $P_{\rm ISCO}$ with each 
other is not unique. It would be equally reasonable to compare Kerr and non-Kerr metrics which 
agree in one or several other observables. 
For example, one can compare models giving the same accretion efficiency \cite{Kong:14}, 
or models giving the same angular offset of the shadow centroid relative to the peak of the 
accretion disk surface brightness. An exhaustive comparison requires sampling 
for each Kerr metric the entire parameter space of alternative metrics.\\[2ex]
 
In the following we assume that $g$ gives the black hole metric as a function of the coordinates
$x^{\mu}\,=\,$ $t,\,r,\,\theta,\,\phi$.  The location of the event horizon of any stationary axially 
symmetric spacetime can only depend on $r$ and $\theta$. 
Assuming that the event horizon is a hypersurface defined by a scalar function $f$ 
and that it is symmetric around the equatorial plane at $\theta=\pi/2$, the null condition 
$\partial_{\mu}f\partial^{\mu}f=0$ reduces in the equatorial plane to 
$g^{rr}=g_{rr}^{-1}=0$ \cite{Thor:07,Joha:13b}.  The event horizon is given 
by the largest $r$ at which the condition holds.

\begin{figure}[t]
\begin{center}
\includegraphics[width=0.45\textwidth]{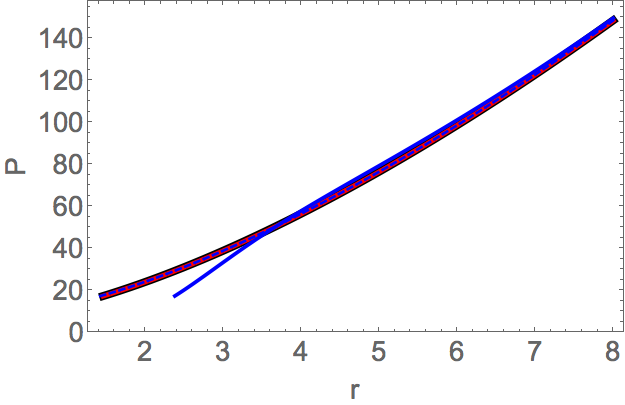}
\end{center}
\caption{Orbital periods as a function of the radial coordinate $r$ for the
Kerr metric with $a=0.99$,
the {\it m2} test metric with $a=$0.872, $\beta=0.2254$
and $a=0.9998$, $\beta=-0.01992$ (all three giving the red solid line), 
and the {\it m6} test metric with
$a=$0.05, $\alpha_{22}=191.08$ (blue solid line) and 
$a=0.999$, $\alpha_{22}=-0.088$ (blue dashed line).
A black hole mass of $M=1$ is assumed and 
$r$ and $P$ are given in geometric units.   
\label{Fperiod}}
\end{figure}

The properties of circular orbits can be derived from the Lagrangian of a test particle. 
The presentation of Equations (\ref{eb1})-(\ref{eb2}) below follows the concise derivation from Bambi et al. (2017) \cite{Bamb:17b}. See~\cite{Cart:68,Chan:83,Ryan:95} for earlier derivations. The Lagrangian is:
\begin{equation}
L=\frac{1}{2}g_{\mu\nu}u^{\mu}u^{\nu}
\label{eb1}
\end{equation} 
with the four velocity $u^{\mu}=\frac{d}{d\tau} x^{\mu}$.
As $L$ does not depend on $t$ and $\phi$, the Euler-Lagrange equations
\begin{equation}
\frac{d}{d\tau}\frac{\partial L}{\partial u^{\mu}}-\frac{\partial L}{\partial x^{\mu}}=0
\end{equation}
lead to two constants of motion, the energy and angular momentum per unit mass:
\begin{equation}
-E\,=\,g_{tt}u^t+g_{t\phi}u^{\phi}\,,\,L_z\,=\,g_{t\phi}u^t+g_{\phi \phi}u^{\phi}.
\label{el}
\end{equation}
Denoting the four velocity of equatorial circular orbits with ${\bf v}$, 
circular orbits require $v^r=\frac{d}{d\tau} v^r=v^{\theta}=0$, so that
the $r$-component of the Euler-Lagrange reads:
\begin{equation}
 (\partial_r  g_{tt})(v^t)^2+2(\partial_r g_{t\phi})v^t v^{\phi}+(\partial_r  g_{\phi\phi})(v^{\phi})^2=0
\end{equation}
After dividing by $v^{\phi}$, solving for the orbital frequency $\Omega=v^{\phi}/v^t$ gives: 
\begin{equation}
\Omega\,=\,(-\partial_r g_{t\phi}\pm\sqrt{(\partial_r g_{t\phi})^2-\partial_r g_{tt} \partial_r g_{\phi\phi}})/\partial_r g_{\phi\phi}.
\label{omega}
\end{equation}
The orbital period of the accreting matter measured by an observer in the 
asymptotically flat region of the spacetime is given by $P=2\pi/\Omega$. 
Figure \ref{Fperiod} shows the $r$-dependence of $P$ for
test particles for the Kerr metric, and {\it m2} and {\it m6} metrics giving the 
same $P$ at the respective ISCOs.

Writing the four velocity of matter on a circular orbits as:
\begin{equation}
v^{\mu}\,=\, v^t (1,0,0,\Omega),
\label{k1}
\end{equation} 
$v^t$ follows from the normalization condition $v^{\,2}=1$:
\begin{equation}
v^t\,=\,(-g_{tt}-2\Omega g_{t\phi}-\Omega^{\,2}g_{\phi\phi})^{-1/2}
\label{k2}
\end{equation}
The constants of motion for equatorial circular orbits follow from the Equations (\ref{el}):
\begin{equation}
E_{\rm CO}\,=\,-(g_{tt}+\Omega g_{t\phi})v^t\,,\,\,
L_{\rm z,CO}\,=\,(g_{t\phi}+\Omega g_{\phi\phi})v^t
\end{equation}

The ISCO is found by solving for (see e.g. \cite{Baum:10}, Sect. 12.1):
\begin{equation}
\frac{dE_{\rm CO}}{dr}=0.
\label{isco}
\end{equation}

Going back to Equations (\ref{el}) for arbitrary orbits, the normalization  condition 
$u^2=1$ gives:
\begin{equation}
g_{tt}(u^r)^2+g_{\phi\phi}(u^{\theta})^2+V_{\rm eff}=1
\end{equation}
with the quasi-potential
\begin{equation}
V_{\rm eff}\,=\,\frac{E^2 g_{\phi\phi}+2 E L_{\rm z} g_{t\phi}+L_{\rm z}^{\,2}g_{tt}}{g_{t\phi}^{\,2}-g_{tt} g_{\phi\phi}}
\label{eb2}
\end{equation}
From the requirements that $u^r=u^{\theta}=\frac{d}{d\tau} u^r=\frac{d}{d\tau} u^{\theta}=0$ for stable circular orbits, 
it follows that $V_{\rm eff}=1$, and $\partial_r V_{\rm eff}=\partial_{\theta} V_{\rm eff}=0$ 
when evaluated in the equatorial plane ($\theta=\pi/2$). Once the ISCO is inferred from Equation (\ref{isco}), 
one can check the radial and vertical stability conditions for all larger orbits. 

One of the pseudo-observables considered below is the redshift of photons coming from 
certain portions of the accretion disk. For simplicity we assume that the photons are emitted 
into a direction vertical to the accretion disk as measured by an observer comoving with the 
emitting plasma with four velocity ${\bf v_{\rm e}}$.
We use a tetrad (an orthogonal set of tangent basis vectors ${\bf e}_{({b})}$ normalized to 
-1, 1, 1, and 1 for the indices ${b}=0,\,1,\,2,$ and 3, respectively) defining a Lorentzian
coordinate frame for the comoving observer: 
\begin{eqnarray}
{\bf e}_{(0)}&\equiv& {\bf v}_{\rm e},  \label{a0}\\
{\bf e}_{(1)}&\equiv & g_{rr}^{\,-1/2}\,\partial_{r},\\
{\bf e}_{(2)}& \equiv & g_{\theta\theta}^{\,-1/2}\,\partial_{\theta},
\label{a1}
\end{eqnarray} 
and  ${\bf e}_{(3)}$ given by the orthonormality conditions.
In terms of these basis vectors, the components of the wave vector 
of photons emitted into the direction $-\partial_{\theta}$ into 
the upper hemisphere are $k_{\rm e}^{\,{b}}=(1,0,-1,0)$. 
The world vector {\bf k}$_{\rm e}$ is thus given by: 
\begin{equation}
{\bf k}_{\rm e}=k_{\rm e}^{\,{b}}e_{({b})} ={\bf e}_{(0)}-{\bf e}_{(2)}.
\label{a2}
\end{equation}
which can be used to read off the contravariant components $k_{\rm e}^{\,\mu}$
in terms of the coordinate basis vectors $(\partial_t, \partial_r,\partial_{\theta},\partial_{\phi})$.
The time translation symmetry of the considered metrics and the associated
Killing vector $\partial_t$ imply that the covariant $t$-component $k_t$ 
of the photon's wave vector {\bf k} is conserved along the photon geodesic 
and keeps its value at the time of emission:
\begin{equation}
k_t=(k_{\rm e})_t=g_{t\mu}k_{\rm e}^{\,\mu}.
\label{a3}
\end{equation}
The fractional frequency change of the photon between emission in the plasma frame 
and the detection by a distant receiver at rest in the asymptotically flat region 
(four velocity {\bf v}$_{\rm r}^{\,\mu}=$ $(1,0,0,0)$) is thus given by 
\begin{equation}
g_{\nu}\,=\,\frac{  {\bf v_{\rm r}\cdot {\bf k}_{\rm r} } }{ {\bf v_{\rm e}\cdot{\bf k}_{\rm e} } }
\,=\, 
\frac{({k}_{\rm r})_t }{-1}\,=\,-\,(k_{\rm e})_t.
\end{equation}
The last equality follows from the conservation of $k_t$ along the geodesic.
 
Page and Thorne (1974) showed that mass, energy, and angular momentum 
conservation together with the assumption of a vanishing torque at the ISCO 
determine the radial brightness profile of the accretion disk \cite{Page:74}. 
The time average flux of radiant energy (energy per unit proper time and unit proper area) 
flowing out of the upper surface of the disk measured by a co-rotating observer is given by:
\begin{equation}
F(r)\,=\,\frac{\dot{M}}{4\pi}e^{-(\nu_0+\psi+\mu_1)} f(r)
\label{EPT2}
\end{equation}
with $\dot{M}$ being the mass accretion rate, and $\nu_0$, $\psi$, and $\mu_1$
from the metric in standard form (Equation (\ref{metric})).
The function $f$ depends on the four velocity of the orbiting particles and its change with
the coordinate $r$:
\begin{equation}
f(r)\,=\,
\frac{-v^t_{\,\,,r}}{v_{\phi}}\,
\int_{r_{\rm ISCO}}^{r}\,
\frac{v_{\phi,r}}{v^t}\,dr
\label{EPT3}
\end{equation}
where ``,'' denotes ordinary partial differentiation.  It is straight forward to solve \eqs{EPT2}{EPT3} numerically.
The bolometric luminosity scales with the mass accretion rate $\dot{M}$ and the accretion efficiency $\eta$.
The latter can be calculated here from the change of a test particle's energy at infinity as it
accretes from infinity to $r_{\rm ISCO}$: $\eta=1-E_{\rm CO}(r_{\rm ISCO})$.  

The disk emits thermally at a radius-dependent temperature of: 
\be
T_{\rm eff}(r)=\left(\frac{F(r)}{\sigma_{\rm SB}}\right)^{1/4}
\label{temp}
\ee
with $\sigma_{\rm SB}$ the Stefan Boltzman constant.
The emitted energy spectrum can be described by a diluted blackbody spectrum 
with a hardening factor of $f_{\rm h}$. The latter parameter gives the blueshift 
of the emitted photons owing to the Comptonization of the photons 
in the accretion disk atmosphere. For stellar mass black holes, the hardening ratio 
has a value between 1.5 and 1.7 \cite{Shim:95,Davi:05,Davi:06}.\\[2ex]

In the following we examine a particular Kerr metric with $M=1$ and $a=0.99$.  
Rapidly spinning black holes are the more interesting ones as the accretion disks
can extend close to the event horizon, where the strong-gravity effects are most pronounced.  
The corresponding event horizon is found at  $r_{\rm H}=1.14$ and the ISCO 
is located at $r_{\rm ISCO}=1.45$.  The orbital angular frequency is $\Omega=0.36$ 
and the period $P$ is $2\pi/\Omega=17.24$. 

We will compare the Kerr metric with several {\it m2} and {\it m6} metrics. 
The metric {\it m2} describes black holes rather than naked singularities for
(with $M=$1): 
\begin{equation}
a^2-\beta^2 \le 1
\label{v1}
\end{equation}
with an event horizon at:
\begin{equation}
r_+=1+\sqrt{1-a^2-\beta^2}
\end{equation}
The deviation parameter $\alpha_{22}$ of Johannsen's test metric {\it m6} modifies the $(t,t)$ and $(t,\phi)$ 
elements of the metric. In the equatorial plane, the event horizon is located at:
\begin{equation}
r_+=1+\sqrt{1-a^2},
\end{equation}
a result formally equal to the one for the Kerr metric. The metric does not exhibit pathologies outside 
of the event horizon as long as \cite{Joha:13}:
\begin{equation}
\alpha_{22}>-(1+\sqrt{1-a^2})^2.
\label{v2}
\end{equation}
Unfortunately, Johannsen's metric yields very unwieldy expressions for most of the quantities of interest
($r_{\rm ISCO}, \Omega, ...$) which we do not reproduce here.  

\begin{figure*}[t]
\begin{center}
\includegraphics[width=0.45\textwidth]{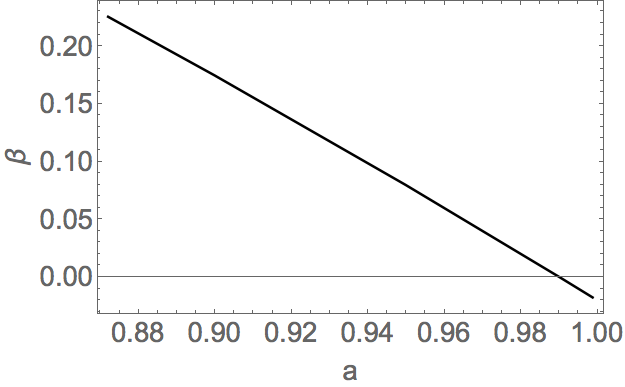}
\includegraphics[width=0.45\textwidth]{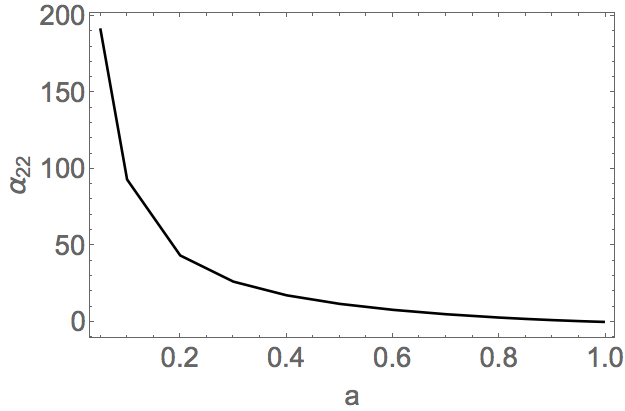}
\end{center}
\caption{\label{F1D} Combinations of the parameters $a$ and $\beta$ of the test metric {\it m2} 
(left) and the parameters $a$ and $\alpha_{22}$ of the test metric {\it m6} (right)
giving the same orbital period at the innermost stable orbit $P_{\rm ISCO}=17.24$
as the Kerr metric with $M=1$ and $a=0.99$.}
\end{figure*}
Setting $M=1$, we neglect the fact that $M$ is poorly constrained 
for many of the systems under study. Possible variations of $M$ 
exacerbate the problem distinguishing between different metrics.
We limit the following discussion to prograde orbits. 
The same analysis could be performed for retrograde orbits.  
The condition $P_{\rm ISCO}=17.24$  defines one dimensional regions in the $a-\beta$ and 
$a-\alpha_{22}$ planes. Figure \ref{F1D} shows these regions as determined 
by a numerical root finder for valid black hole metrics satisfying Equation (\ref{v1}) or Equation (\ref{v2}). 
Curves such as the ones shown in Fig.~\ref{F1D} establish maps of the original parameters $(a,\beta)$ 
and $(a,\alpha_{22})$ to the new parameters ($P_{\rm ISCO},a$) for the overlapping $P_{\rm ISCO}$ regions. 
For the Kerr metric $P_{\rm ISCO}$ can take all values between $4\pi$ and $52\pi$.
The fact that the {\it m2} and {\it m6} metrics can produce $P_{\rm ISCO}$-values outside of this range proves criterion {\it C2} that the two test metrics include spacetimes which are physically 
distinct from any Kerr spacetime. 
Note that observational tests in the region of the {\it m2} and {\t m6} parameter space with $P_{\rm ISCO}>52\pi$ face additional problems as disk instabilities might move the inner edge of the disk to $r$-values $r_1$ exceeding $r_{\rm ISCO}$ with an associated orbital period $P(r_1)\gg P_{\rm ISCO}$. The problem can be mitigated somewhat by repeated observations of one and the same object enabling the determination of $r_{\rm ISCO}$ (and/or $P_{\rm ISCO}$) as the lower bound(s) of the observed values.
Criterion {\it C2} can further be proven by showing that the different metrics produce non-identical observables
for one and the same $P_{\rm ISCO}$-value. We will do so in the next paragraph.

In the following we will look at a few observables, comparing the Kerr metric with $a=0.99$, with the 
nearly-extreme {\it m2} metrics with $a=$0.872, $\beta=0.2254$ and $a=0.9998$, $\beta=-0.01992$ 
and the nearly extreme {\it m6} metrics with $a=$0.05, $\alpha_{22}=191.08$ and $a=0.999$, $\alpha_{22}=-0.088$. All metrics give the same $P_{\rm ISCO}=17.24$. 

\begin{figure}[t]
\begin{center}
\includegraphics[width=0.45\textwidth]{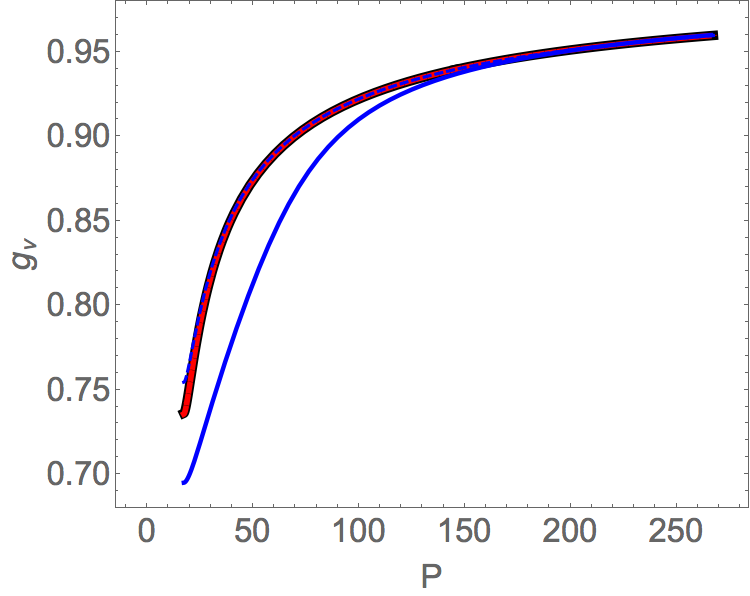}
\end{center}
\caption{\label{Fgfac} Net fractional change $g_{\nu}$ of the frequency of vertically (plasma frame) 
emitted photons reaching an observer at infinity as a function of the orbital period 
for the Kerr metric with $a=0.99$ (black solid line, shadowed by the red line), 
the {\it m2} test metric with $a=$0.872, $\beta=0.2254$ (red solid line)
and $a=0.9998$, $\beta=-0.01992$ (red dashed line, shadowed by the red line), 
and the {\it m6} test metric with
$a=$0.05, $\alpha_{22}=191.08$ (blue solid line) and 
$a=0.999$, $\alpha_{22}=-0.088$ (blue dashed line).}
\end{figure}
Figure \ref{Fgfac} shows the fractional frequency change $g_{\nu}$ of vertically emitted photons
as a function of the orbital period $P$.
Whereas the two {\it m2} metrics and the {\it m6} metric with $a=0.999$ and $\alpha_{22}=-0.088$ 
give the same $g_{\nu}$-values, the {\it m6} metric with $a=0.05$ and $\alpha_{22}=191.08$ gives noticeably smaller 
$g_{\nu}$-values (higher redshifts) for $P<130$ ($r<{\sim} 7 r_{\rm g}$).
The figure indicates that the observations of Fe K$\alpha$ emission from matter spiraling towards 
the black hole can be used to distinguish between a black hole described
by the Kerr metric one of the non-Kerr {\it m6} metrics. 

Using the equations described above, we can estimate the radial distribution of the 
temperature $T$ and brightness $F$ of the accretion disk photosphere. 
\begin{figure}[t]
\begin{center}
\includegraphics[width=0.45\textwidth]{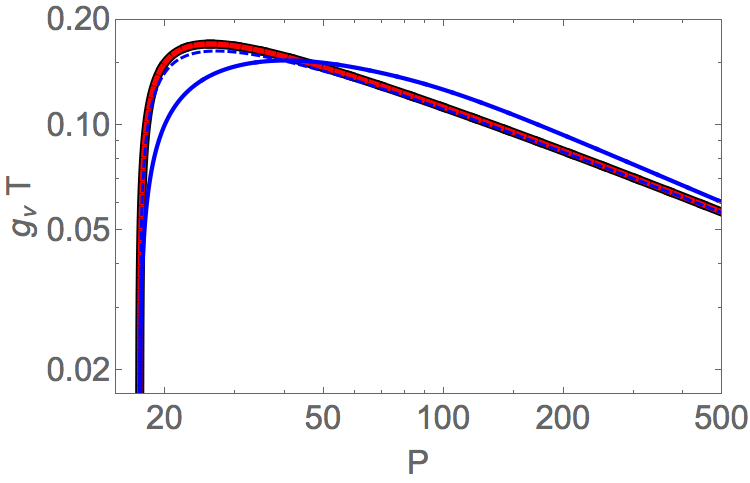}
\end{center}
\caption{\label{Fgt} Product of the photospheric accretion disk temperature 
times the frequency shift $g_{\nu}$ between emission and observer 
as a function of the orbital period for the same metrics as in Fig.\ \ref{Fgfac}.}
\end{figure}
Figure \ref{Fgt} shows the product of $T$ and the redshift factor $g_{\nu}$ as a proxy for the energies of the
observed photons as a function of $P$.  Interestingly, the results for the Kerr metric, the two {\it m2} metrics, and the {\it m6} metric with $a=0.999$ and $\alpha_{22}=-0.088$ are again indistinguishable. Only the {\it m6} metric with $a=0.05$ and $\alpha_{22}=191.08$ gives a different distribution with - compared to the distributions for the Kerr metric - lower temperatures closer to the black hole 
and higher temperatures further away. 
 
\begin{figure}[t]
\begin{center}
\includegraphics[width=0.45\textwidth]{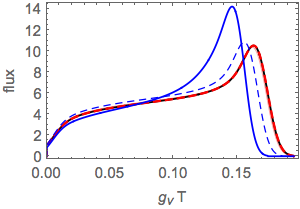}\newline
\includegraphics[width=0.45\textwidth]{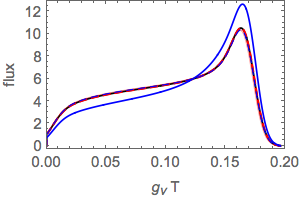}
\end{center}
\caption{\label{Fes} Distributions of the accretion disk temperatures weighted with the observer frame
energy flux for identical accretion rates (top) and after fine tuning the accretion rate and the flux normalization to make the distributions as similar as possible (bottom) for the same metrics as in Fig.\ \ref{Fgfac}.}
\end{figure}

Assuming an accretion disk annulus at radius $r$ emits continuum emission at observed energy 
$E_{\gamma}\propto g_{\nu}\,T$, we can generate pseudo energy spectra by histogramming the
temperature of all accretion disk ring annulli with $r\in \left[ r_{\rm ISCO}, \infty\right]$ 
weighing each annulus with the energy flux of the annulus (Figure \ref{Fes}, upper panel). 
As before, only the {\it m6} metric with $a=0.05$ and $\alpha_{22}=191.08$ gives noticeable deviations. 
The lower panel of Figure \ref{Fes} shows that these deviations persist when fine tuning 
the mass accretion rate $\dot{M}$ and the flux normalization to minimize the difference between the distributions.

To summarize this section: the best strategy for comparing the spacetimes of different metrics is to 
focus exclusively on observables. Approximate degeneracies between metric parameters can be 
eliminated by replacing one or more of the degenerate parameters with key observables.
Our analysis indicates that the metric {\it m2} is not well suited for X-ray tests of the Kerr hypothesis
as it barely impacts X-ray observables. In contrast, the metric {\it m6} can be used as it 
does lead to quite different observable outcomes. A full ray-tracing simulation should 
be used for a more thorough evaluation.
\section{Constraining black hole spacetimes with X-ray observations}
\label{Xobs}
Two methods for measuring the spins of black holes have been extensively used:
(i) fitting the continuum energy spectra of stellar mass black holes in the thermal state, and
(ii) modeling the energy spectra from the inner accretion disk, including the 
Fe K$\alpha$ fluorescent line, and, if observationally accessible, the Compton hump emission.
Whereas the first method can only be used for stellar mass black holes in X-ray binaries, 
the second method can be used for mass accreting stellar mass and supermassive black holes. 

The {\bf thermal continuum fitting} method (e.g.\ \cite{Zhan:97,Gier:01,McCl:14})
is based on observing the thermal emission of geometrically thin, optically thick accretion disks.
The method requires the independent measurement of 
the distance $D$ of the X-ray binary, 
the black hole mass $M$, and 
the inclination $i$ of the binary system.
The inclination $i$ is defined as the angle between the angular momentum axis of the binary 
and the line of sight. These quantities can be inferred from radio, infrared and/or optical
observations of the binary system. 
Assuming that the disk is described by the standard Novikov-Thorne model, 
and that the angular momentum vectors of the black hole and the binary 
system are aligned, the X-ray energy spectrum can then be used to fit the 
black hole spin parameter $a$ (determining the ISCO) and the accretion rate $\dot{M}$ using suitable emission models. 
State-of-the-art models include relativistic effects (frame dragging, Doppler and gravitational frequency shifts, 
light bending) and detailed modeling of the emission including limb darkening and spectral hardening, and 
disk self-irradiation.

The {\bf inner disk reflection modeling} method \cite{Fabi:89,Bren:06,Mill:07,Bren:13} 
relies entirely on modeling the X-ray energy spectra. The reflected emission is thought to originate 
from the irradiation of the accretion disk photosphere with
hard coronal X-ray emission. The gravitational and Doppler frequency shifts broaden the reflected lines, 
and more so for high spins (small $r_{\rm ISCO}$) and large inclinations $i$. 

In the following we discuss the results for stellar mass black holes in X-ray binaries as the spins obtained 
with both methods can be compared to each other. Furthermore, the inclinations inferred from the inner disk 
reflection modeling can be compared to the non-X-ray constraints on the inclination of the binary orbit.   
Bambi, Jiang and Steiner (2017) give a compilation of recent continuum fitting and inner disk reflection modeling results. The list includes 19 stellar mass black hole candidates and 25 
supermassive black holes \cite{Bamb:17} (see also \cite{Bren:13,Fabi:16}).

\begin{figure*}[t]
\begin{center}
\includegraphics[width=0.45\textwidth]{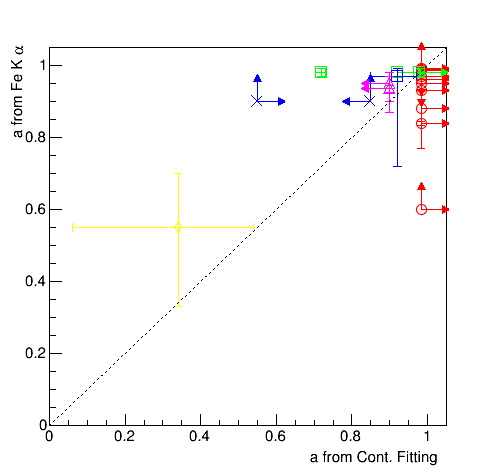}
\includegraphics[width=0.45\textwidth]{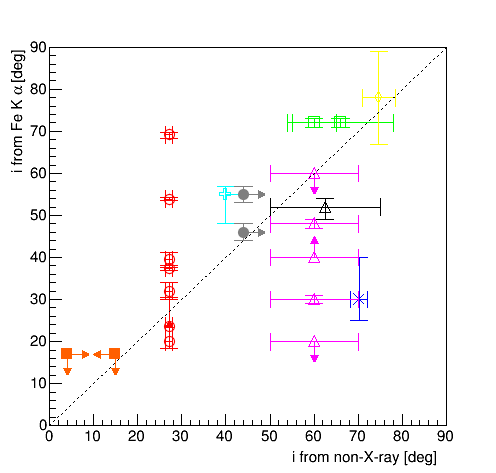}
\end{center}
\caption{\label{Fdata} The left panel compares the black hole spin parameter estimates 
from the continuum fitting method and the inner reflection modeling for the stellar mass
black holes 
Cyg X-1 (red open circles) \cite{Gou:14,Duro:11,Fabi:12,Mill:12,Toms:14,Walt:16,Toms:18},
GRO 1655-40 (blue crosses) \cite{Shaf:06,Reis:09},
GRS 1915+105 (open green squares) \cite{McCl:06,Midd:06,Mill:13,Reid:14},
GX 339$-$4 (open magenta triangles) \cite{Kole:10,Reis:08,Garc:15,Park:16},
LMC X-1 (blue open squares) \cite{Gou:09,Stei:12a},
XTE J1550$-$564 (yellow diamond) \cite{Stei:11,Stei:12}.
The right panel compares the inclinations from non-X-ray observations and
inner reflection modeling for the black holes
Cyg X-1 (red open circles) \cite{Oros:11,Fabi:12,Duro:11,Walt:16,Toms:14,Toms:18},
GRO 1655-40 (blue cross) \cite{Gree:01,Reis:09},
GRS 1915+105 (open green squares) \cite{Fend:99,Mill:13,Reid:14},
GX 339$-$4 (open magenta triangles) \cite{Kole:10,Reis:08,Fuer:15,Garc:15,Park:16},
MAXI J1836-194 (filled orange squares) \cite{Russ:14,Reis:12},
SWIFT J1753.5-0127 (open cyan cross) \cite{Neus:14,Reis:09},
V404 Cygni (black open triangle) \cite{Khar:10,Walt:17}, 
XTE J1550$-$564 (yellow diamond) \cite{Stei:11,Stei:12a}, and
XTE J1650-500 (grey filled circles) \cite{Oros:04,Mill:09,Walt:12}.
}
\end{figure*}

The spin parameters for all six stellar mass black holes with results from both methods are 
shown in the left panel of Fig. \ref{Fdata}.
The confidence intervals are all from the original publications and are usually on the 67\% or 90\% confidence level 
for measured values and on the 99\% or 3$\sigma$ confidence level for lower or upper limits.
A rough agreement between the results obtained with the two methods can be recognized in the sense that
most of the studied black holes have rather high spin parameters exceeding 0.5. 
For a few of the objects the two methods yield only marginal agreement or even significant disagreement:
\begin{description}
\item [Cyg X-1:] For Cyg X-1, we have a highly reliable distance estimate of
1.86$_{-0.11}^{+0.12}$ kpc from the measurement of a trigonometric parallax with the 
Very Long Baseline Array \cite{Reid:11} and good constraints on the mass of the companion 
$M_*=(19.2 \pm 1.9)M_{\odot}$, the mass of the black hole $M_{\rm BH}=(14.8 \pm 1.0)M_{\odot}$, 
and the inclination of the binary orbit
$i = 27.1^{\circ} \pm 0.8^{\circ}$ \cite{Oros:11}. 
The spin parameter result $a\, >\, 0.983$ (3$\sigma$) from the 
thermal continuum method \cite{Gou:14} is thus particularly well supported by observations.
The inner disk reflection modeling gives a wide range of different results:    
$0.6\le a \le 0.99$ \cite{Mill:12}, 
$a=0.838\pm0.006$ \cite{Toms:14},
$a=0.88_{-0.11}^{+0.07}$ \cite{Duro:11},
$0.93\le a \le 0.96$ \cite{Walt:16},
$a=0.949_{-0.019}^{+0.013}$ \cite{Toms:18},
$a=0.97_{-0.02}^{+0.014}$ \cite{Fabi:12}, 
$a=0.9882_{-0.009}^{+0.009}$ \cite{Toms:14}, and
 $a>0.987$ \cite{Toms:18}.
\item [GRO 1655$-$40:] Whereas the thermal continuum fitting method gives $0.55 \le a \le 0.85$ \cite{Shaf:06},
the  inner disk reflection modeling gives $a>0.9$ \cite{Reis:09}.  The discrepancy seems to be significant as
both groups go through a number of models with all thermal continuum fitting results $<$0.85 and all
inner disk reflection modeling results $>0.9$.    
\item [GRS 1915+105:] McClintock et al. (2006) present a thermal continuum fitting analysis 
indicating $a>0.98$ \cite{McCl:06}. The authors give a very detailed evaluation of the impact of various 
observational uncertainties on the results, showing that substantially lower spin parameters ($a\sim 0.8$) 
are less likely but possible. Middleton et al. (2006) get a significantly lower thermal continuum 
result of $a=0.72_{-0.017}^{+0.009}$  \cite{Midd:06}, a discrepancy which might result from using
intermediate rather than low luminosity observations for the analysis \cite{McCl:06}.
Reid et al.\ (2014) derive another continuum fitting result of $a>0.92$. 
Miller et al.'s (2013) inner disk reflection modeling 
indicates a near-extremal spin parameter of $a=0.98_{-0.01}^{+0.01}$ \cite{Mill:13}.
\item[GX 339$-$4:] Kolehmainen \&Done (2011) set a limit of $a<0.9$ on the 
spin parameter of GX 339$-$4 from the analysis of the thermal continuum emission \cite{Kole:11}
which is lower than the inner disk reflection modeling results of 
$a=0.935$ $\pm$~0.01 (statistical) $\pm$~0.01 (systematic) by Reis et al.~\cite{Reis:08},
$a=0.95_{-0.05}^{+0.03}$ by Garc\'ia et al. \cite{Garc:15}, and
$a=0.95_{-0.08}^{+0.02}$ by Parker et al. \cite{Park:16}. 
\end{description}
The discussion shows the systematic errors are still large ($\Delta a \sim 0.1-0.2$), 
as evident from the different results obtained with one and the same method
(Cyg X-1 and GRS 1905+105), and with the two complimentary methods (GRO 1655$-$40 and GX 339$-$4).\\[2ex]

Similarly interesting is the comparison of the orbital inclination and the inclination of the inner disk 
from the inner disk reflection modeling. The right panel of Fig.~\ref{Fdata} shows consistent results for 
GRS 1915+105, SWIFT J1753.5-0127, V404 Cygni, XTE J1550-564, 
XTE J1650-500, and MAXI J1836-194. For Cyg X-1, GRO 1655$-$40, and GX 339$-$4 
there are significant deviations:
\begin{description}
\item [Cyg X-1:] The modeling of the optical photometric and spectroscopic data indicates a near face-on inclination of
$i=27.1^{\circ} \pm 0.8^{\circ}$ \cite{Oros:11}. 
The inner disk reflection modeling gives a wide range of inclinations 
$i<20^{\circ}$ \cite{Toms:18}, 
$i=23.7^{\circ}-5.4^{\circ}+6.7^{\circ}$ \cite{Fabi:12},
$i=32^{\circ}\pm 2^{\circ}$ \cite{Duro:11},
$i=37.5^{\circ}\pm0.7^{\circ}$ \cite{Toms:18},
$37.6^{\circ}\le i \le 41.3^{\circ}$ \cite{Walt:16},
$i=53.9^{\circ}\pm 0.4^{\circ}$ and 
$i=69.2^{\circ}-0.9^{\circ}+0.5^{\circ}$ \cite{Toms:14}.
\item [GRO 1655$-$40:] The inclination $i=70.2^{\circ}\pm 1.9^{\circ}$ \cite{Gree:01} 
deviates significantly from the inner disk reflection modeling result of $i=30^{\circ}-5^{\circ}+10^{\circ}$ \cite{Reis:09}.
\item [GX 339$-$4:] Kolehmainen \& Done (2010) argue for a likely orbital inclination of
between 50$^{\circ}$ and 70$^{\circ}$ \cite{Kole:10}.
The inner disk reflection modeling gives a range of results: 
$i<20^{\circ}$ \cite{Reis:08},
$i=40^{\circ}-60^{\circ}$ \cite{Fuer:15},
$i=30^{\circ}\pm 1^{\circ}$ \cite{Park:16}, and
$i=48^{\circ}\pm 1^{\circ}$ \cite{Garc:15}.
\end{description}

The discrepancies could result from a misalignment of the inner disk and the binary orbit 
(see  \cite{Macc:02,Mart:08}). 
If the angular momentum vectors of the black hole and the binary system are not aligned a Bardeen-Petterson
type configuration may result with the inner accretion disk orbiting the black hole in its equatorial plane and the
outer disk being aligned with the binary orbit (see the discussion in Sect.\ \ref{disc}).  

How would the thermal continuum black hole spin parameters change if one would use the inclinations 
from the inner disk reflection modeling in the analysis?
For Cyg X-1 the thermal continuum fitting spin parameter would drop from close to 1 to $a \sim 0.96$ 
when changing the inclination from $i\approx 20^{\circ}$ to $i \approx 40^{\circ}$ (\cite{Gou:11}, Fig. 5) 
making it consistent with the spin parameter ($0.93\le a \le 0.96$) and inclination ($37.6^{\circ}\le i \le 41.3^{\circ}$) 
results of Walton et al. (2016) \cite{Walt:16}. The high inclination of $i= 69.2^{\circ}$ from the 
inner disk reflection model \#4 of Tomsick et al. (2014) would bring the thermal 
continuum fitting spin parameter down to $a\sim 0.9$, significantly lower than the corresponding 
inner disk reflection spin parameter of $a=0.9882-_{0.009}^{0.009}$.
Using the inclination of $i= 53.9^{\circ}$ from the inner disk reflection model \#8 of Tomsick's paper
would bring the thermal continuum fitting spin parameter down to $a\sim 0.93$, still significantly higher
than the spin parameter of $a=0.838\pm0.006$ from the inner disk reflection modeling.
For GRO 1655$-$40, using the inclination $i\approx 30^{\circ}$ from the inner disk reflection 
modeling instead of the orbital inclination of $i\approx 70^{\circ}$ would resolve the discrepancy
between the thermal continuum fitting results  ($a<$0.85) and the inner disk reflection 
modeling ($a>0.9$). The misalignment hypothesis can thus resolve some but not all of the
discrepancies. Note that Steiner et al.~constrain the jet-disk misalignment to be less than $12^{\circ}$ 
for one particular source (XTE J1550$-$564) \cite{Stei:12}.

If we adopt the hypothesis that the inner accretion disks of some stellar mass black holes are misaligned,
the main concern evident from Fig. \ref{Fdata} are the widely different results obtained 
with one and the same method for the well studied objects Cyg X-1 and GRS 1915+105.
McClintock et al. (2006) explain the difference between their and Middleton et al.'s thermal continuum fitting results
by the use of different data sets. They argue that the observations of low-luminosity ($L_{\rm X}<0.3 L_{\rm Edd}$) 
rather than intermediate-luminosity thermal state observations give the most reliable results. 
At higher luminosities the disk may acquire a non-negligible thickness, and the relative importance 
of a non-vanishing torque at $r_{\rm ISCO}$ may increase. Analyzing a large number of low-luminosity
data sets, McClintock et al. (2006,2014) \cite{McCl:06,McCl:14} find excellent agreement 
for all the individual results.
Noble et al. (2011) \cite{Nobl:11}, Kulkarni et al. (2011) \cite{Kulk:11}, and Zhu et al. (2012) \cite{Zhu:12} 
study the impact of a non-vanishing inner torque on the radial emission profile and the inferred 
thermal continuum fitting spin parameter results based on GRMHD simulations. They find that the associated
systematic errors are rather small, especially for high spins 
(e.g. $\Delta a \sim 0.3, 0.1, 0.03, 0.01$ for spin parameters of $a<0.5, a=0.7, 0.9,$ and 0.98, respectively).
One major theoretical uncertainty of the thermal continuum fitting method is the spectral hardening 
by a factor $\sim f_{\rm h}$ impacting the mean energy of thermally emitted photons 
$E_{\gamma}\approx 2.70 f_{\rm h} k_{\rm B} T$ where $k_{\rm B}$ is the Boltzmann constant. 
For example, Shafee et al. (2006) \cite{Shaf:06} 
find that the thermal continuum spin parameters of the two black holes GRO J1655-40 and 4U 1543-47 ($a\sim 0.7$) 
change by about $\Delta a\approx $0.1 between the 
spectral hardening models of Shimura \& Takahara (1995) \cite{Shim:95} and 
Davis et al. (2005) \cite{Davi:05}.

The {\bf inner disk reflection modeling} depends on disentangling the continuum emission, 
the reflection component, and absorption features (see e.g. \cite{Kole:11,Bren:13}). A broad bandpass 
as that afforded by NASA's {\it NuSTAR} mission helps to do so (e.g.\ \cite{Risa:13}). 
The fit models depend on assumptions about the geometry and physical properties of the accretion disk, 
the accretion disk photosphere, and the corona (or the emissivity profile). Usually, the accretion disk is assumed
to extend from $r_{\rm ISCO}$ to infinity and the corona is assumed to be a point source of  (in its rest frame) isotropic emission hovering above the black hole. Alternatively, a certain functional form of the emission profile 
(i.e.~the radial dependence of the coronal flux impinging on the accretion disk, prompting 
the emission of Fe K$\alpha$ photons) is assumed (e.g. a single power law or a broken power law) 
and the analysis includes fitting a number of parameters describing the emissivity profile. 
The line shapes and equivalent widths are assumed to depend only on the metallicity of the plasma, 
and the ionization parameter. These assumptions may not be correct:
\begin{itemize}
\item The material in the plunging region between the event horizon and the inner edge of the accretion disk
may modify the observed line shapes (however, see the discussions in \cite{Mill:07}).
\item Accretion disks may have an inner cutoff at $r_1>r_{\rm ISCO}$, they may have non-negligible thicknesses, 
and they may be warped or clumpy.
\item The ionization degree and density of the photospheric plasma may vary strongly as a function
of the radial coordinate \cite{Garc:16}.
\item The coronas may have different and/or time varying geometries, see \cite{Daus:13,Garc:14,Gonz:17}.
\item The emission from some parts of the accretion disks may be absorbed.
\end{itemize}

Various authors noted that some of the observed Fe~K$\alpha$ lines requires unrealistically high metallicities exceeding solar metallicities by factors as large as ten or higher, indicating that the inner reflection line modeling is still missing important physics (see \cite{Reyn:97,Garc:14} and references therein). Garc\'{i}a et al.\ 2016 show that the density of the photospheric plasma strongly impact the
shapes of the reflected lines \cite{Garc:16}. Tomsick et al. (2018) fit {\it NuSTAR} and {\it Suzaku}
observations of Cyg~X-1 and find that a higher-density model gives substantially different 
metallicity, spin, and inclination results \cite{Toms:18}. 
   
Figure  \ref{Fdata} and the systematic errors and uncertainties described above indicate that 
the X-ray constraints on black hole spin parameters and inclinations as well as 
X-ray tests of the Kerr hypothesis have to be received with some caution.
We will discuss possible avenues for further progress in the next section.  
\section{Discussion}
\label{disc}
According to the Kerr hypothesis, astrophysical quasi-stationary black holes are macroscopic 
elementary particles can be described by four continuous parameters (mass, angular momentum
magnitude, and angular momentum orientation). As quasi-stationary black holes play important roles 
in galaxies and galaxy clusters, and describe key objects involved in gravitational wave events, 
it is highly desirable to test this prediction as accurately as possible.
X-ray tests are independent and complementary to tests based on gravitational waves, 
radio interferometric observations of black hole shadows, and the observations of stars
orbiting supermassive black holes.  
The discussion above can be summarized as follows:
\begin{enumerate}
\item The Kerr family of metrics describes a wide range of physically different spacetimes. 
Many of the proposed alternative metrics produce very similar spacetimes.
 \item Solar system and binary tests of GR are based on observations of 
 isolated test bodies in stable orbits, enabling precision measurement of the properties 
 of the underlaying spacetime.  In the case of X-ray observations of black holes, 
 we see the emission from accretion disks: viscous, self-interacting, non-linear, macroscopic objects. 
Small differences of the underlying spacetime are easily drowned by the averaging 
over different orbits effected by the turbulence of the accreting plasma. 
The macroscopic properties of the disk are then largely determined by energy, mass, 
and angular momentum conservation.
\item Although the numerical modeling of black hole accretion has made enormous progress 
over the last two decades, the observational outcomes depend on the proper modeling of many 
different physical processes. The observational outcomes may depend on the detailed properties
of the magnetized, partially ionized plasma, and the detailed phase space distributions 
of photons and electrons. We may still be far away from a sufficiently complete understanding 
of  which properties and processes play an important role and which ones may be neglected. 
\item Even if the numerical simulations captured all the relevant physical processes, 
the problem of mapping the observations to a certain accretion flow configurations 
may be ill defined. It may simply not be possible to use the observations to constrain 
all the relevant properties of the accretion flow (e.g. the shape and location of the corona, 
the alignment of the black hole spin and the angular momentum vector of the accretion disk,
and the properties of the accreted magnetic field) sufficiently well to allow for quantitative 
tests of the Kerr hypothesis. 
\item The analysis of the X-ray observations are plagued by several practical challenges 
contributing uncertainties to the derived results, including the selection of suitable data sets 
and the accurate modeling of the contributions from other emission components.   
\end{enumerate}
The spin parameter and inclination results discussed in Sect. \ref{Xobs} indicate that the current uncertainties are rather large. For example, the spin parameter seems to be uncertain by 
$\Delta a\sim 0.1-0.2$. The finding emphasizes the difficulties one faces when using the 
X-ray data for quantitative tests of GR. 

Several upcoming missions and missions in development 
can add new information about the inner accreting flows:

{\it X-ray polarimetry:} The {\it Imaging X-ray Polarimetry Explorer (IXPE)} \cite{Weis:16}, 
NASA's first dedicated X-ray polarimetry mission is scheduled for launch in 2021 and will measure 
the polarization of X-ray sources in the 2-8 keV energy band. 
The mission will acquire high signal-to-noise observations of bright stellar mass black holes 
in X-ray binaries and first polarimetric results for a number of nearby Seyfert 1 galaxies. 
The polarization of the thermal continuum emission of mass accreting stellar mass black holes
will give new constraints on the orientation of the angular momentum vector of the inner accretion disk 
and improved measurements of the  black  hole spin parameter \cite{Li:09,Schn:09,Kraw:12}. 
{\it IXPE} and balloon borne experiments like {\it X-Calibur} \cite{Kisl:17} will 
test hypotheses about the physical properties of the corona of accreting 
stellar mass and supermassive black holes
\cite{Schn:10,Schn:13g,Kraw:16,Behe:17} and the nature of the reflected 
emission \cite{Dovc:04,Mair:18}.  
We expect that X-ray polarization observations will allow us to validate (or falsify) the current 
accretion disk and corona models in a similar way as broadband X-ray observations of
stellar mass and supermassive black holes with {\it NuSTAR} have tested the 
reflection nature of the Fe K$\alpha$ line emission (e.g. \cite{Risa:13}).

{\it High-throughput X-ray spectroscopy:} Future high-throughput X-ray spectroscopy missions 
such as the upcoming Advanced Telescope for High-ENergy Astrophysics ({\it ATHENA}) mission of the 
European Space Agency (anticipated launch: 2028) \cite{Nand:11}, the Spectroscopic Time-Resolving Observatory for Broadband Energy X-rays ({\it STROBE-X}) \cite{Wils:17},  or the Enhanced Timing and X-ray Polarimetry mission ({\it eXTP}) of the Chinese Academy of Sciences \cite{Zhan:16} can perform time resolved 
studies of the Fe K$\alpha$ emission. 
As mentioned above, sensitive observations of frequency shifts 
(owing to gravitational and Doppler frequency shifts) as a function of the distance 
from the black hole (encoded in the orbital modulations of the signal) 
might be able to distinguish between Kerr and non-Kerr spacetimes (see also \cite{Boll:15}).

{\it Black Hole Imager:} Ultimately, we would like to image accretion disks.  
Currently, the Event Horizon Telescope (EHT) combines the 230~GHz data from several 
Very Long Baseline Interferometry (VLBI) telescopes around the Earth.
The expected angular resolution of between 15 and 20 $\mu$arcsec of the EHT is 
comparable to the angular extents of the supermassive black holes at the centers of 
Sgr A$^*$ and M~87 (e.g. \cite{Akij:17}). The first results are expected to be announced soon. 
Similar angular resolutions might be obtained in the X-ray regime in the more distant future 
using interferometric techniques or transmissive, refractive-diffractive optics \cite{BHI}.
The radio and X-ray observations would be impacted by very different astrophysical and instrumental
systematics. Obtaining images with both techniques would thus be highly desirable.\\[2ex]

Continued observations with {\it Chandra}, {\it XMM-Newton}, {\it NICER}, and {\it NuSTAR} 
as well as observations with future experiments are likely to open up new ways of testing the Kerr hypothesis:

{\it Observations of microlensed quasars:} 
The X-ray observations of some gravitationally lensed quasars show evidence for microlensing by stars.
The amplitude distribution of the flux variations indicate that the X-ray bright regions 
(i.e. the X-ray emitting coronas) of the microlensed quasars are smaller than 
$\sim 30 r_{\rm g}$ \cite{Char:09,Dai:10,Morg:12,Mosq:13,Blac:15,MacL:15}.
For some of the lensed sources, the energy spectra show evidence for the relativistically 
broadened Fe K$\alpha$  emission \cite{Char:12,Reis:14,Char:17}. 
The {\it Chandra} observations of the lensed quasars RX J1131$-$1231, 
SDSS~1004+4112, QJ~0158$-$4325, and MG J0414+0534
reveal a shift of the line centroid for some of the observations, 
and multiple lines for others \cite{Char:12,Char:17,2002ApJ...568..509C}. 
A possible explanation for the observations is that the line shifts are caused by the 
selective amplification of the emission from certain regions of the accretion disk 
as the microlensing caustics move across the system \cite{2002ApJ...568..509C,Char:17,Kraw:17}. 
Based on this paradigm, Chartas et al. (2017) use the energies of the
detected lines to constrain the inclination and spin parameter of RX J1131$-$1231. 
The observations offer a variant on the inner disk reflection modeling, and share many of the 
same systematics. In particular, the analysis of the observational data relies on 
the modeling of the Fe K$\alpha$ emission. The amplification of the X-rays from 
different regions of the accretion disk and the corona do add new information 
about the inner accretion flow. However, the results also depend on a number 
of additional parameters, i.e.\ the  convergence and shear of the macrolens 
and the location, orientation, and scale of the microlensing caustic(s).
Acquiring observations over many years from many different  caustic crossings, 
as well as dense observational sampling of a single caustic crossing might 
make it possible to overconstrain the quasar and lens parameters.

\begin{figure}[t]
\begin{center}
\includegraphics[width=0.45\textwidth]{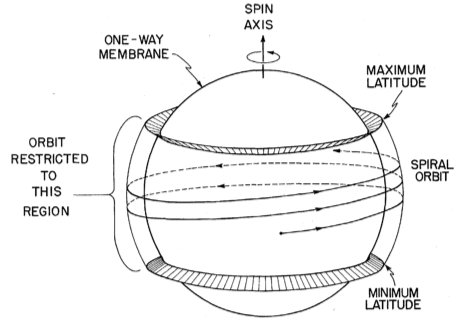}
\end{center}
\caption{\label{Fwk} Part of an inclined test particle orbit around a maximally rotating Kerr black hole showing 
a corkscrew pattern close to the event horizon rather than the nodal advance for Lense-Thirring 
precession of orbits further away from the black hole (from \cite{Wilk:72}).}
\end{figure}
{\it Observations of Relativistic Precession:} 
The thin disk theory described above applies for systems with parallel black hole and
accretion disk angular momentum vectors. Misaligned systems can show much richer 
dynamics owing to the interplay of relativistic frame dragging 
and the turbulence of the accreting plasma. 
The effect of frame dragging on test particles orbiting a spinning body of mass $M$ 
and angular momentum $a$ is commonly referred to as Lense-Thirring precession of the orbits. 
Far away from the spinning body, the frame dragging advances the nodes of circular orbits by an angle 
$\Delta \Omega= 2 (a/M) (M/r)^{3/2}$ per revolution \cite{Lens:18}. 
Close to a black hole, the frame dragging effects are so strong 
that test particles orbit on corkscrew-type trajectories (Fig. \ref{Fwk}) \cite{Wilk:72}. 
The combination of the precessing orbits and the turbulence
may produce a Bardeen-Petterson-type configuration 
with the inner disk orbiting the black hole in the equatorial 
plane of the black hole out to a characteristic radius beyond 
which the disk maintains its original misalignment \cite{Bard:75,Papa:83,Kuma:85,Zhur:11}.  
Alternatively, warps may propagate with the speed of sound giving rise to bending 
waves with a periodically changing disk inclination at a 
given distance from the black hole \cite{Papa:95,Demi:97,Ivan:97,Lubo:02}.
The relativistic precession of the inner disk may cause the low-frequency 
quasi-periodic oscillations (LFQPOs, 0.1-30 Hz) of the X-ray fluxes from
accreting neutron star and black hole sources (e.g.\ \cite{Stel:98,Stel:99,Schn:06,Ingr:09,Vele:13,Stef:14}).
Ingram and Done (2009) \cite{Ingr:09} speculate that the inner disk precesses like a solid object. 
Based on this paradigm, Ingram et al. (2015, 2017) \cite{Ingr:15,Ingr:17} explain the
quasi-periodic variations of the energy of the reflected Fe K$\alpha$ line in GRS 1915+105 and 
H~1743$-$322 as the result of the reflection of the coronal emission 
off a precessing inner accretion disk. The net precession depends on the underlying 
spacetime through the range of available stable orbits and the properties of these orbits. 
The confrontation of semi-analytical models of misaligned accretion systems 
with the results from GRMHD simulation (see \cite{Frag:07,Frag:09,Zhur:14,Mora:14}) 
will lead to a better understanding of the relevant physical processes.
X-ray polarimetry will give us new observational constraints on the orientation of the inner accretion disks 
and can be used for sensitive searches for such precession effects \cite{Ingr:15}.

{\it Observations of Quasi-Periodic Oscillations (QPOs):} The light curves of several 
black holes in low-mass X-ray binaries (LMXBs) exhibit QPOs identified as peaks in the 
power density spectra (PDS) \cite{Remi:06,vand:06,Zhan:13}. 
If high-frequency quasi-periodic oscillations (HFQPOs, 40-450 Hz) were 
directly related to the frequencies of test particle orbits 
(i.e.~the orbital frequency, the radial and vertical epicyclic frequencies, 
and the Lense-Thirring precession frequency), 
QPOs could be powerful tools for constraining the underlying spacetime  
(e.g. \cite{Abra:01,Abra:05}). 
It seems more likely however, that the QPO frequencies 
result from a complex interplay of the orbital kinematics and the highly non-linear 
properties of the accretion disk plasma, making the observed frequencies and implications 
for the underlying spacetime highly model dependent. Intriguingly, QPO's are only found in 
hard state light curves, not in the thermal state light curves. The finding indicates that 
the geometrically thin and optically thick accretion disks responsible for the 
thermal state emission reach down to the ISCO and do not precess. 
 
Confronting the observational results from ongoing and upcoming X-ray missions 
with higher fidelity numerical models promises to give us new insights into the 
physics of black hole accretion, and should eventually give us more reliable constraints 
on the properties of the underlying background spacetimes.
\begin{acknowledgements}
I thank Q. Abarr, B. Beheshtipour, P.~Bolt, M. Errando, C. Gammie, J. Garc\'{i}a, 
B. Groebe, A. Ingram, F. Kislat, and J. Miller for highly enjoyable and helpful discussions. 
I am grateful to the anonymous referees whose excellent comments improved the paper 
substantially. I acknowledge NASA funding through the awards 
80NSSC18K0264 and NNX16AC42G. 
\end{acknowledgements}

\end{document}